\documentclass[aps,twocolumn]{revtex4}
\usepackage{graphicx}
\usepackage{dcolumn}

\begin{document}

\title{Dynamical real-space renormalization group calculations
with a new clustering scheme on random networks}
\author{D. Balcan$^{1}$ and A. Erzan$^{1,2}$}
\affiliation{$^1$ Department of Physics, Faculty of Sciences and
Letters\\
Istanbul Technical University, Maslak 34469, Istanbul, Turkey \\
$^2$ G\"ursey Institute, P.O.B. 6, \c Cengelk\"oy, 34680 Istanbul,
Turkey }

\date{\today}

\begin{abstract}
We have defined a new type of
clustering scheme preserving the connectivity of the nodes in
network ignored by the conventional Migdal-Kadanoff bond moving
process. Our new clustering scheme performs much better for
correlation length and dynamical critical exponents in high
dimensions, where the conventional Migdal-Kadanoff bond moving
scheme breaks down. In two and three dimensions we find the
dynamical critical exponents for the kinetic Ising Model  to be $z=2.13$ and $z=2.09$,
respectively at pure Ising fixed point. These values are in very
good agreement with recent Monte Carlo results. We investigate the phase diagram
and the critical behaviour for randomly bond diluted lattices in d=2 and 3, in
the light of this new transformation. We also provide exact correlation exponent
and dynamical critical exponent values on hierarchical lattices with power-law
degree distributions, both in the pure and random cases.


\end{abstract}

\maketitle
\section{Introduction}
We have generalized the dynamical real-space renormalization group
(RSRG) calculations for the kinetic Ising model~\cite{Suzuki} to
dilute lattices with arbitrary number of nearest neighbors,
motivated by an interest in the relaxation behaviour on
networks~\cite{Dorogovtsev} with scale free and exponential degree
distributions.

\begin{figure}
\leavevmode
\rotatebox{0}{\scalebox{.7}{\includegraphics{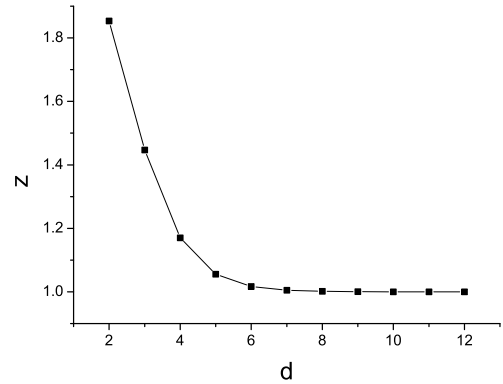}}}
\caption{The dynamical critical exponent $z$ versus dimension $d$
obtained via the conventional Migdal-Kadanoff bond moving scheme.
$z$ converges to 1 for large $d$.} \label{dynexpMK}
\end{figure}

\begin{figure}
\leavevmode
\rotatebox{0}{\scalebox{.7}{\includegraphics{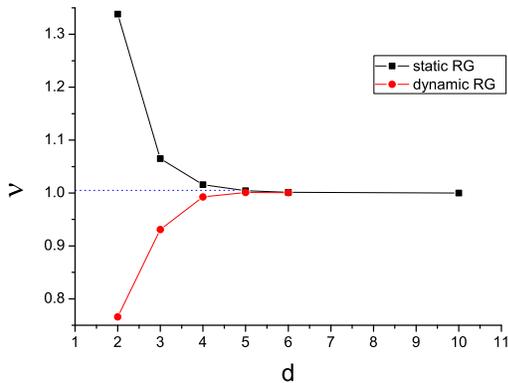}}}
\caption{The correlation critical exponent $\nu$ versus dimension
$d$ via the conventional Migdal-Kadanoff bond moving scheme. $\nu$
converges to 1 for large $d$.} \label{correxpMK}
\end{figure}

We first computed the dynamical critical exponent $z$, in the
Migdal-Kadanoff bond moving scheme~\cite{Migdal,Kadanoff} on
networks with arbitrarily high, but uniform, connectivity and
found as have others~\cite{Droz,DrozMalaspinas1}, that $z$
gradually converges to unity, as the spatial dimension of the
system becomes very large (for $d=12$, $z-1=10^{-5}$), and the
correlation length exponent converges to $\nu=1$. (See Fig.
\ref{dynexpMK} and \ref{correxpMK}) This  is in contrast to the
expected Mean Field values of $\nu=0.5$ and $z=2$ above the upper
critical dimension. The dynamic real-space renormalization group
calculation thus yields neither a sharp crossover above the upper
critical dimension, $d_{\rm c}$=4, nor the correct Mean Field
behaviour in the high dimension limit. We have observed that a
static RSRG calculation with Migdal-Kadanoff bond moving scheme (MK)
also converges to the same limits, but it does so from above,
whereas the dynamical RSRG calculation does so from below. (See
Fig. \ref{correxpMK})

\begin{figure}
\leavevmode
\rotatebox{0}{\scalebox{.7}{\includegraphics{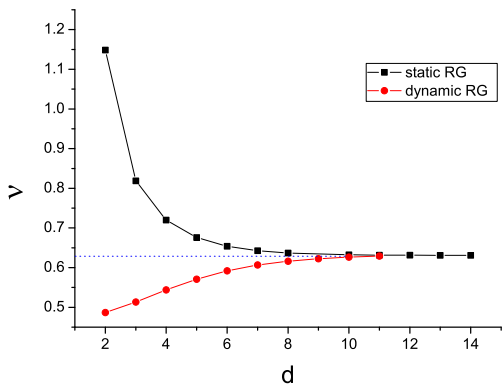}}}
\caption{The correlation critical exponent $\nu$ versus dimension
$d$ obtained by our new clustering scheme. $\nu$ converges to 0.63
for large $d$ (compare it with Fig. \ref{correxpMK}).}
\label{correxp}
\end{figure}

\begin{figure}
\leavevmode
\rotatebox{0}{\scalebox{.7}{\includegraphics{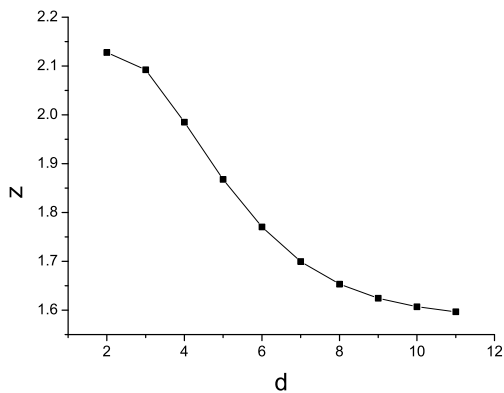}}}
\caption{The dynamical critical exponent $z$ versus dimension $d$
obtained by our new clustering scheme. $z$ converges to 1.6 for
large $d$ (compare it with Fig. \ref{dynexpMK}).} \label{dynexp}
\end{figure}

The reason why the MK approach fails to provide a reasonable approximation to
the critical behaviour of $d$ dimensional hypercubic lattices in large $d$, is
twofold. The first is because it underestimates the contribution
from the loops in the $d$-dimensional Euclidean lattice and this
effect leads to more and more inaccurate results for large
dimensions. We have found a way to improve the performance of the
RS approach in relation to $d$-dimensional hypercubic lattices, by
defining a new type of RSRG cluster, which retains the
inter-connectivity of the moved spins, and were able to obtain a
convergence to $\nu=0.63$ (better than 1 but still larger than
0.5) in the limit of large dimensionality ($d>13$). ( See Fig.
\ref{correxp}) Likewise, the dynamical critical exponent in this
scheme converges to the value 1.6 for dimensions $d \ge 11$. (See
Fig. \ref{dynexp})
The new clustering scheme for the RSRG also performs very well in
low dimensions. The dynamical exponent calculated within our
scheme for $d=2$ is $z=2.13$ and for $d=3$ is $z=2.09$, to be
compared with the best simulation
results~\cite{Stauffer,Nightingale,Li,Lauritsen,Adler,ItoOzeki,Ito},
which give values between 2.11 and 2.24 for $d=2$ and between 2.01
and 2.11 for $d=3$.

The second reason for the divergence of MK results from those on Euclidean
hypercubic lattices, is the nonuniform topology of the underlying hierarchical
lattice, on which the MK scheme for RSRG is realised as an exact
transformation.~\cite{Berker,Kaufman} We turn this feature to an advantage in
investigating the dynamical behaviour of the Ising model on networks with
power-law  degree distributions.

Going over to the dynamical critical behaviour for bond diluted lattices, we
first applied the
dynamic RSRG technique to the bond diluted kinetic Ising system in
two and three dimensions. The approximation we used to compute the
configuration averages is in fact exact for annealed randomness.
The approximate transformations for the dilution parameter $p$ and
the coupling constant are well behaved as one approaches the
separatrix, or critical line, from above, but the approximation
breaks down in the ordered phase and for low temperatures in the
disordered phase.
In both two and three dimensions we were able to compute the RG
flows on the disordered side of the separatrix and on the
separatrix itself, and thereby determine the phase diagrams. In
two dimensions we find a disorder fixed point, which is unstable,
and which we interpret as a tricritical point, since there the
second order phase transition line gives way to a first order
transition line. The pure Ising fixed point is stable, and
determines the exponents along the second order transition line.
In three dimensions, no disordered critical fixed point is found.
The critical line is depressed to zero temperature at a
concentration $p_e > p^*$, where $p^*$ is the percolation fixed
point. The flow along the critical line is to the pure Ising fixed
point at $p=1$, and thus the critical exponents along the critical
line are the same as the pure Ising exponents, also in three
dimensions.
Computing the effective critical exponents along the critical
line, we find that $z_{\rm eff}$ varies non-monotonically as a
function of $p$, within the intervals $[2.01, 2.25]$ for $d=2$ and
$[2.09, 2.69]$ for $d=3$.

Our
scheme as well as the conventional equilibrium Migdal-Kadanoff RSRG (see also
Ref. ~\cite{DrozMalaspinas2,Andelman}) fails to
predict the crossover to a disorder critical fixed point for $d=3$, both
demonstrated by means of finite size scaling arguments~\cite{Domany} applied to
large Monte Carlo simulations~\cite{Ballestros} and expected on the basis of the
Harris criterion~\cite{Harris}.  The value we find  for the pure system specific
heat exponent $\alpha$ via the hyper-scaling realtion $2- \alpha= d \nu$  in
$d=3$ is negative for the static RSRG calculation for $\nu$, while the dynamic
calculation yields a positive $\alpha$.
These results have to be interpreted in the context of the still ongoing debate
on the criteria for the stability of the pure-system critical behaviour.  The
rather extensive literature on the Harris criterion~\cite{Harris} has been
recently reviewed by Janke and Weigel~\cite{Janke}.
It has been shown by various authors~\cite{Andelman,Luck,Efrat} that the Harris
criterion, which equates the crossover exponent for randomness, $\phi$ to 
the pure system $\alpha$, is simply not applicable on hierarchical lattices, and
various alternative criteria, such as the ``wandering exponent''~\cite{Luck} for
correlations in the non-periodic variations in the number of bonds incident on
lattice points (the degree of the node) have been proposed. The calculation of
this exponent for our present RSRG scheme goes beyond the scope of this paper
and will be considered in a separate publication.

The paper is organized as follows. In the next section we setup
the dynamical RSRG calculations for bond diluted hypercubic
lattices, and introduce a new clustering scheme on which we will
implement it. The last section includes our results and a
discussion of the relevance of our results to non-uniform lattices
with power law degree distributions.

\section{The dynamical RSRG calculations for bond diluted hypercubic lattices}
In order to investigate the effect of the underlying lattice of
arbitrarily high degree on dynamical behaviour of an interacting
system living on this lattice, we consider an Ising model on the
nodes of a hypercubic lattice of $d$-dimensions, which will be
subjected to bond dilution to yield a random network with a Poisson degree distribution.

\begin{figure}
\leavevmode
\rotatebox{0}{\scalebox{.55}{\includegraphics{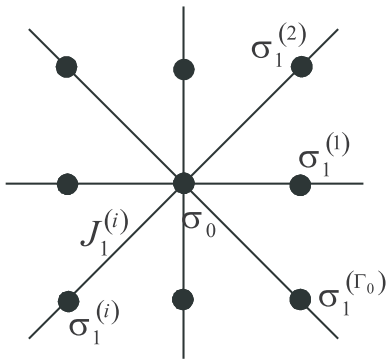}}}
\caption{The``central'' spin on which we concentrate is denoted by
$\sigma_0$, its nearest neighbor spins by $\sigma_{1}^{(i)}$, and
the interaction constants by $J_{1}^{(i)}$, where $i$ takes values
from $1$ to $\Gamma_0$.} \label{notation}
\end{figure}

The hamiltonian of the system is given by
\begin{equation}
H = - \sum_{\langle ij \rangle} J_{ij} \; \sigma_i \; \sigma_j \;,
\end{equation}
where $J_{ij}$ is the interaction between two nearest neighbor
spins and $\sigma_i$ is the spin variable which can take the
values $+1$ and $-1$. The sum is taken over all nearest neighbor
pairs. In a $d$-dimensional hypercubic lattice the maximum number
of nearest neighbor spins with a nonzero interaction is $\Gamma_0
= 2d$. Now we will concentrate on a spin and its neighborhood: we
will denote this spin by $\sigma_0$, the nearest neighbor spins
and the interaction constants by $\sigma_{1}^{(i)}$ and
$J_{1}^{(i)}$, respectively, where $i$ takes values from $1$ to
$\Gamma_0$ as shown in Fig. \ref{notation}. Here $J_{1}^{(i)}$ is
subjected to the distribution,
\begin{equation}
\label{j-dist}
P(J_{1}^{(i)}) = p \delta(J_{1}^{(i)}-J) + (1-p)
\delta(J_{1}^{(i)}) \;.
\end{equation}

\subsection{Equation of motion for the magnetization}
Using Glauber dynamics~\cite{Glauber} we may write down the
equation of motion for the magnetization $m_0 \equiv \langle
\sigma_0 \rangle $ and get
\begin{equation}
{d \over dt} m_0(t) = -m_0(t) + \langle \tanh \left(
\sum_{i=1}^{\Gamma_0} K_{1}^{(i)} \sigma_{1}^{(i)} \right) \rangle
\label{corner0} \;,
\end{equation}
where $K_{1}^{(i)} \equiv \beta J_{1}^{(i)}$ and $\beta \equiv
(k_BT)^{-1}$. Here the brackets $\langle \; \rangle$ denote both
the thermal and the configuration averages. If we let $\Gamma$ be
the number of nearest neighbors interacting with a spin
$\sigma_0$, then the distribution of $\Gamma$ is given by
\begin{equation}
P(\Gamma) = \sum_{n=0}^{\Gamma_0} {\Gamma_0 \choose n} (1-p)^n
p^{\Gamma_0-n} \delta(\Gamma-(\Gamma_0-n)) \;.
\end{equation}
Now if we expand the function appearing inside the brackets in
Eq. (\ref{corner0}), in terms of spin products, and then sum over
all possible configurations, we obtain
\begin{eqnarray}
\left( 1 + {d \over dt} \right) m_0(t) &=& \left[ {
\sum_{\Gamma=1}^{\Gamma_0} {\Gamma_0-1 \choose \Gamma-1}
p^{\Gamma} (1-p)^{\Gamma_0-\Gamma} a_{\Gamma}(K) } \right]
\nonumber \\
&& \times \sum_{i=1}^{\Gamma_0} m_{1}^{(i)} + g(p,K,t)
\label{corner00} \;.
\end{eqnarray}
Here $a_{\Gamma}(K)$ is the coefficient of the fist order term
coming from the expansion of the function in terms of products of
spin variables, for a particular configuration in which spin
$\sigma_0$ has $\Gamma$ interacting nearest neighbors. These
coefficients are given by
\begin{equation}
a_{\Gamma}(K) = { 1 \over \Gamma \; 2^{ \Gamma - 1 } }
\sum_{n=0}^{n_{\rm max}} { \Gamma \choose n } ( \Gamma - 2 n )
\tanh[(\Gamma - 2 n)K] \;,
\end{equation}
where $n_{\rm max}$ takes values $\Gamma/2$ for even $\Gamma$ or
$(\Gamma-1)/2$ in the case of odd $\Gamma$. The other term in Eq.
(\ref{corner00}), $g(p,K,t)$ comes from the higher order spin
products. We see that we can write the equation of motion for a
spin variable as
\begin{equation}
\left( 1 + {d \over dt} \right) m_0(t) = a(p,K)
\sum_{i=1}^{\Gamma_0} m_{1}^{(i)} \label{corner1} \;,
\end{equation}
where we ignore the higher order terms and define $a(p,K)$ as
\begin{equation}
a(p,K) = \sum_{\Gamma=1}^{\Gamma_0} {\Gamma_0-1 \choose \Gamma-1}
p^{\Gamma} (1-p)^{\Gamma_0-\Gamma} a_{\Gamma}(K)
\label{coefficient1} \;.
\end{equation}
If we take the Laplace transform of the equation (\ref{corner1})
we obtain the equation of motion for the magnetization of a spin
as
\begin{equation}
\left( 1 + s \right) m_0[s] = a(p,K) \sum_{i=1}^{\Gamma_0}
m_{1}^{(i)} \label{laplace1} \;.
\end{equation}

\subsection{New clustering scheme}
\begin{figure}
\leavevmode
\rotatebox{0}{\scalebox{.45}{\includegraphics{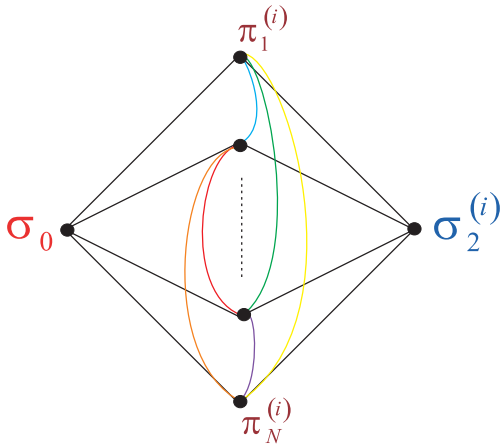}}}
\caption{The cluster in our new scheme which will, in the coarse
grained lattice, go to the bond connecting the central spin to its
nearest neighbor in the $i$th direction. The intermediate
(``middle") spins are denoted by $\pi_{j}^{(i)}$, $j=1, \ldots,
N$. This approach preserves the inter-connectivity of the
``middle" spins and is thus able to estimate better the number of
different paths, or loops, contributing to the spin correlations.
(Our figure is in color in the on-line version.)} \label{cluster}
\end{figure}

In this subsection we will introduce a new type of clustering
scheme other than the conventional Migdal-Kadanoff bond moving
approach. Since Migdal-Kadanoff bond moving drastically
underestimates the number of loops in the system in high
dimensions, we need to introduce a new scheme preserving the
inter-connectivity of the nodes which will be eventually decimated
(see Fig. \ref{cluster}). We will implement the transformation for
the scaling parameter $\lambda=2$.

According to our new scheme, in the $i$th direction,
$i=1,\ldots,\Gamma_0$, between the ``corner" spins $\sigma_0$ and
$\sigma_{2}^{(i)}$ there exists a cluster containing
$N=\lambda^{d-1}$ ``middle" spins denoted by $\pi_{j}^{(i)}$, with
$j=1,2,\ldots,N$, as illustrated in Fig. \ref{cluster}. All the
``middle" spins can be connected to the ``corner" spins and to the
rest of the ``middle" spins in the cluster. The ``corner" spins
will remain after decimation, and may have a maximum number
$\Gamma_{0,\rm c} = \Gamma_0 N$ of nearest neighbors. The
``middle" spins will be decimated through the last step of the
dynamical RG calculations, and the maximum number of nearest
neighbors they may have is $\Gamma_{0, \rm m}= N+1$. We postulate
that the bond occupation probability $p$, and the bond strength
$J$ remain invariant after the application of our new type of bond
moving scheme, and the interaction between any pair of spins on
the cluster obeys the same distribution as in Eq. (2).
 The
distribution of the number of nearest neighbors is given by
\begin{equation}
P(\Gamma') = \sum_{n=0}^{\Gamma'_{\rm max}} {\Gamma'_{\rm max}
\choose n} (1-p)^n p^{\Gamma'_{\rm max}-n}
\delta(\Gamma'-(\Gamma'_{\rm max}-n)) \;,
\end{equation}
where $\Gamma'_{\rm max}$ is $\Gamma_{0, \rm c}$ for the ``corner"
spins and $\Gamma_{0, \rm m}$ for the ``middle" spins.

The equation of motion for the expectation value of the $j$th
``middle" spin in the $i$th direction now becomes,
\begin{equation}
\left( 1 + {d \over dt} \right) \langle \pi_{j}^{(i)} \rangle =
\gamma(p,K) \left[ m_0 + m_{2}^{(i)} + \sum_{k \not= j} \langle
\pi_{k}^{(i)} \rangle \right] \label{middle1} \;,
\end{equation}
where $\gamma(p,K)$ comes from the configuration average given by
\begin{equation}
\gamma(p,K) = \sum_{\Gamma'=1}^{\Gamma_{0,\rm m}} {\Gamma_{0,\rm
m}-1 \choose \Gamma'-1} p^{\Gamma'} (1-p)^{\Gamma_{0,\rm
m}-\Gamma'} a_{\Gamma'}(K) \;.
\end{equation}
If we write down the equation of motion for the expectation value
of the ``corner" spins we obtain
\begin{equation}
\left( 1 + {d \over dt} \right) m_0(t) = A(p,K)
\sum_{i=1}^{\Gamma_0} \sum_{j=1}^{N} \langle \pi_{j}^{(i)} \rangle
\label{corner2} \;,
\end{equation}
where $A(p,K)$ comes from the configuration average,
\begin{equation}
A(p,K) = \sum_{\Gamma'=1}^{\Gamma_{0,\rm c}} {\Gamma_{0,\rm c}-1
\choose \Gamma'-1} p^{\Gamma'} (1-p)^{\Gamma_{0,\rm c}-\Gamma'}
a_{\Gamma'}(K) \;.
\end{equation}

\subsection{Obtaining the dynamical RSRG equations}
Now we are ready to perform the decimation. The aim is to rewrite
the equation of motion for the ``corner" spins in terms of
their $\lambda$th neighbors. In our case, with $\lambda =2$, this means obtaining the equation of
motion for $m_0$ in terms of the  $m_{2}^{(i)}$'s. For this purpose we
will write down all the equations obtained for the ``middle" spins 
$\pi_j^{(i)}$
in the $i$th direction and sum all the equations. We get
\begin{eqnarray}
\left[ 1 + {d \over dt} - (N-1) \gamma(p,K) \right] \sum_{j=1}^{N}
\langle \pi_{j}^{(i)} \rangle = \nonumber \\ N \gamma(p,K) \left[
m_0 + m_{2}^{(i)} \right] \label{middle2} \;.
\end{eqnarray}
Thus we  obtain the equation of motion for the ``middle" spins in
the $i$th direction in terms of the ``corner" spin averages $m_0$
and $m_{2}^{(i)}$. If we multiply the equation of motion
(\ref{corner2}) for $m_0$ by $\left[1 + {d \over dt} - (N-1)
\gamma(p,K)\right]$ we obtain
\begin{eqnarray}
\left[ 1 + {d \over dt} - (N-1) \gamma(p,K) \right] \left( 1 + {d
\over dt} \right) m_0(t) = \nonumber \\ A(p,K)
\sum_{i=1}^{\Gamma_0} \left[ 1 + {d \over dt} - (N-1) \gamma(p,K)
\right] \sum_{j=1}^{N} \langle \pi_{j}^{(i)} \rangle \;.
\end{eqnarray}
Using Eq. (\ref{middle2}) we get
\begin{eqnarray}
\left[ 1 + {d \over dt} - (N-1) \gamma(p,K) \right] \left( 1 + {d
\over dt} \right) m_0(t) = \nonumber \\ A(p,K)
\sum_{i=1}^{\Gamma_0} N \gamma(p,K) \left[ m_0 + m_{2}^{(i)}
\right] \;
\end{eqnarray}
and
\begin{eqnarray}
\left[ \left( 1 + {d \over dt} \right)^2 - (N-1) \gamma(p,K)
\left( 1 + {d \over dt} \right) - \nonumber \right.
\\ \left. N \Gamma_0 A(p,K) \gamma(p,K) \right] m_0(t) = \nonumber
\\ N A(p,K) \gamma(p,K) \sum_{i=1}^{\Gamma_0} m_{2}^{(i)} \;.
\end{eqnarray}
If we ignore the second derivative term we obtain
\begin{eqnarray}
\{ \left[ 1 - (N-1) \gamma(p,K) - N \Gamma_0 A(p,K) \gamma(p,K)
\right] + \nonumber \\ \left[ 2 - (N-1) \gamma(p,K) \right] {d
\over dt} \} m_0(t) = \nonumber \\ N A(p,K) \gamma(p,K)
\sum_{i=1}^{\Gamma_0} m_{2}^{(i)} \;.
\end{eqnarray}
Now let us write this equation in a familiar form,
\begin{eqnarray}
\{ 1 + {2 - (N-1) \gamma(p,K) \over 1 - \gamma(p,K) \left[ N-1 + N
\Gamma_0 A(p,K) \right]} {d \over dt} \} m_0 = \nonumber \\ {N
A(p,K) \gamma(p,K) \over 1 - \gamma(p,K) \left[ N-1 + N \Gamma_0
A(p,K) \right]} \sum_{i=1}^{\Gamma_0} m_{2}^{(i)} \label{corner3}
\;.
\end{eqnarray}
Taking the Laplace transform we get
\begin{eqnarray}
\{ 1 + {2 - (N-1) \gamma(p,K) \over 1 - \gamma(p,K) \left[ N-1 + N
\Gamma_0 A(p,K) \right]} s \} m_0[s] = \nonumber \\ {N A(p,K)
\gamma(p,K) \over 1 - \gamma(p,K) \left[ N-1 + N \Gamma_0 A(p,K)
\right]} \sum_{i=1}^{\Gamma_0} m_{2}^{(i)} \label{laplace2} \;.
\end{eqnarray}

We see that the equation of motion (\ref{laplace2}) is in the same
form as Eq. (\ref{laplace1}). We identify the second term in the
curly brackets as the renormalized Laplace variable $\tilde s$.
The coefficient in front of the summation appearing on the
right-hand side, we identify as the coefficient $a(\tilde p,
\tilde K)$, expressed in terms of the renormalized variables
$\tilde p$, $\tilde K$. We thus obtain the RG equation for the
time from
\begin{equation}
{\tilde s \over s} = {2 - (N-1) \gamma(p,K) \over 1 - \gamma(p,K)
\left[ N-1 + N \Gamma_0 A(p,K) \right]} \;\label{sfix}
\end{equation}
and the implicit RG transformation for $K$ from
\begin{equation}
a(\tilde p, \tilde K) = {N A(p,K) \gamma(p,K) \over 1 -
\gamma(p,K) \left[ N-1 + N \Gamma_0 A(p,K) \right]} \equiv R(p,K)
\;,\label{Kfix}
\end{equation}
where $a(p,K)$ is given by Eq. (\ref{coefficient1}). The
transformation for the renormalized occupation probability $\tilde
p$ is found by calculating the probability $f(p)$ of an unbroken path
from the spin $\sigma_0$ to $\sigma_{2}^{(i)}$ through the
cluster in the $i$th direction, and is thus independent from $K$.
Thus, the fixed point value for the occupation probability
satisfies
\begin{equation}
p^\ast = f(p^\ast) \;.\label{pfix}
\end{equation}
Note that this implies that at each stage of the distribution, the distribution
of the bond strengths is replaced by a distribution of the initial binary form,
Eq.(\ref{j-dist}), with the renormalised parameters $\tilde{p}$ and
$\tilde{K}$.~\cite{Stinchcombe}  This may hide from view certain features of the
random fixed point associated with the full distribution.~\cite{Andelman,Domany}

The fixed point of the dynamical RG transformation for $K$ is
found from
\begin{equation}
a(p^\ast,K^\ast) = R(p^\ast,K^\ast) \;,
\end{equation}
where $p^*$, found from Eq. (\ref{pfix}), should be substituted.
We can evaluate the correlation critical exponent $\nu$ from
\begin{equation}
{d \tilde K \over d K} {\vert}_{p^\ast, K^\ast} = \left[ {\partial
R(p,K) \over \partial K} / {\partial a(\tilde p, \tilde K) \over
\partial \tilde K} \right] {\vert}_{p^\ast, K^\ast} =
\lambda^{-\nu} \;,
\end{equation}
and the dynamical critical exponent $z$ is given by
\begin{equation}
{\tilde s \over s} {\vert}_{p^\ast, K^\ast} = \lambda^z \;.
\label{eqdynexp}
\end{equation}

\section{Results and discussion}

In the foregoing we have presented the generalized dynamical RSRG
framework for the kinetic Ising model on dilted $d$-dimensional lattices with a random number of nearest neighbors, motivated by an interest in the
scaling behaviour of relaxation times on random networks.

For the case with no bond dilution, we calculated the correlation
length critical exponent $\nu$ and dynamical critical exponent $z$
with our new scheme of clustering in $d$-dimensional hypercubic
lattices and found that $\nu$ converges to 0.63 and $z$ converges
to 1.6 for large $d$ values as shown in
Figs.~(\ref{correxp},\ref{dynexp}). The numerical values are given
in Table 1. These results are much better than those obtained by
the conventional Migdal-Kadanoff bond moving scheme which is shown
in Figs. (\ref{correxpMK}, \ref{dynexpMK}), respectively. Note
that, here, just as for the conventional Migdal-Kadanoff bond
moving scheme, these results for the critical point and the
correlation exponent differ from those found directly from the
fixed point of a RSRG transformation obtained by decimating the
middle spins in the cluster shown in Fig.~\ref{cluster}. We denote
the latter scheme as the ``static approach" and have reported the
results for the correlation length exponent found in this way
under $\nu_{\rm static}$.

\begin{table}
\begin{ruledtabular}
\caption{Our results for dynamical critical exponent $z$ and
correlation critical exponent $\nu$ with respect to space
dimension $d$ at the pure Ising fixed point.}
\begin{tabular}{c|c|c|c}
{\boldmath $d$} & {\boldmath $z$} & {\boldmath $\nu_{\rm
dynamic}$} & {\boldmath
$\nu_{\rm static}$} \\
\hline\hline
2  & 2.13 & 0.49 & 1.15 \\
3  & 2.09 & 0.51 & 0.82 \\
4  & 1.99 & 0.54 & 0.72 \\
5  & 1.88 & 0.57 & 0.68 \\
6  & 1.77 & 0.59 & 0.65 \\
7  & 1.70 & 0.61 & 0.64 \\
8  & 1.65 & 0.62 & 0.64 \\
11 & 1.60 & 0.63 & 0.63 \\
\end{tabular}
\end{ruledtabular}
\end{table}

Our new scheme yields dynamical critical exponent values $z=2.13$
and $z=2.09$ in two and three dimensions, as well as the
percolation exponent $\nu_p$. We report our results in Table 2. We
see that the agreement between the known value of $\nu_p=4/3$ in
two dimensions is about as good as the result in three dimensions,
with the best Monte Carlo values being reported as $\nu_p= 0.88$
~\cite{StaufferAharony}. The values of the dynamical critical
exponents are in very good agreement with recent Monte Carlo
results, as summarized in Table 3.

\begin{table}
\begin{ruledtabular}
\caption{The fixed points and the critical exponents in $d=2$ and
$d=3$. The first value of $p$ shows the pure Ising fixed point,
the second one shows the percolation fixed point for each
dimension $d$.}
\begin{tabular}{c|c|c|c|c|c|c}
{\boldmath $d$} & {\boldmath $p$} & {\boldmath $K^\ast$} &
{\boldmath $\nu_p$} & {\boldmath $\nu_{\rm dynamic}$} & {\boldmath
$\alpha_{\rm dynamic}$} &
{\boldmath $z$}\\
\hline\hline
2  & 1     & 0.27 & --   & 0.49 & 1.03 & 2.13  \\
   & 0.5   & 0.82 & 1.43 & 0.48 & 1.04 & 2.16  \\
\hline
3  & 1     & 0.12 & --   & 0.51 & 0.46  & 2.09 \\
   & 0.16  & --   & 1.01 & --   & -- & -- \\
\end{tabular}
\end{ruledtabular}
\end{table}

\begin{table}
\begin{ruledtabular}
\caption{Comparison of our results with the known values coming
from different approaches. Here dynamical RSRG calculations are
written as DRSRG and Monte Carlo studies are denoted by MC.}
\begin{tabular}{c|c|c|c}
$d$ & Reference & Method & {\boldmath $z$} \\
\hline\hline
2  & Present work                                   & DRSRG  & 2.13  \\
   & Stauffer ~\cite{Stauffer}                      & MC     & 2.18  \\
   & Nightingale {\it et al.} ~\cite{Nightingale}   & MC     & 2.17  \\
   & Li {\it et al.} ~\cite{Li}                     & MC     & 2.13  \\
   & Lauritsen {\it et al.} ~\cite{Lauritsen}       & MC     & $2.13 \pm 0.02$
\\
   & Adler {\it et al.} ~\cite{Adler}               & MC     & $2.21 \pm 0.03$
\\
   & Droz {\it et al.} ~\cite{DrozMalaspinas1}       & DRSRG  & 1.85  \\
\hline
3  & Present work                                   & DRSRG  & 2.09  \\
   & Ito {\it et al.} ~\cite{ItoOzeki}              & MC     & 2.06  \\
   & Stauffer ~\cite{Stauffer}                      & MC     & 2.04  \\
   & Lauritsen {\it et al.} ~\cite{Lauritsen}       & MC     & $2.04 \pm 0.03$
\\
   & Adler {\it et al.} ~\cite{Adler}               & MC     & $2.08 \pm 0.03$
\\
   & Ito ~\cite{Ito}                                & MC     & $2.06 \pm 0.02$
\\
   & Droz {\it et al.} ~\cite{DrozMalaspinas1}       & DRSRG  & 1.45  \\
\hline
6  & Present work                                   & DRSRG  & 1.77  \\
   & Droz {\it et al.} ~\cite{DrozMalaspinas1}       & DRSRG  & 1.02  \\
\end{tabular}
\end{ruledtabular}
\end{table}

\begin{figure}
\leavevmode
\rotatebox{0}{\scalebox{.75}{\includegraphics{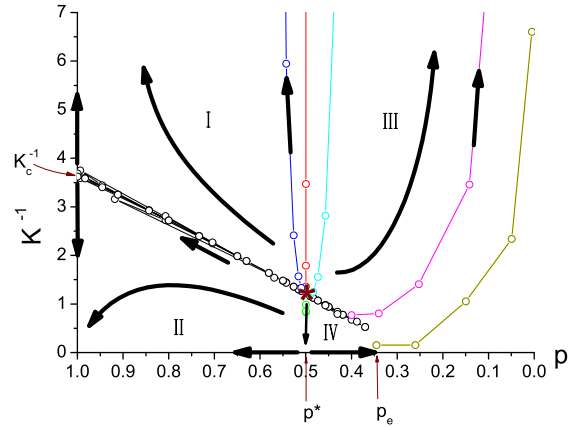}}}
\caption{The phase diagram, $K^{-1} = k_BT/J$ versus $p$, for
$d=2$. There is an unstable fixed point at $p^{\ast}=0.5$,
$({K^{\ast}})^{-1}=1.22$, indicated by $\ast$ in the figure. The
line of points extending to the right of the unstable fixed point
is explained in the text. The flow from region IV is to the high
temperature, $p=0$ fixed point. Thus the line from ($p^{\ast},
T^{\ast}$) to the percolation fixed point, $(p^\ast, 0)$ is a
first order phase transition line. The critical behaviour on the
phase boundary extending from $\ast$ to the pure Ising fixed point
at $T_{\rm c}$ at $p=1$, is determined by the latter point. (Our
figure is in color in the on-line version.)} \label{2d}
\end{figure}

\begin{figure}
\leavevmode
\rotatebox{0}{\scalebox{.8}{\includegraphics{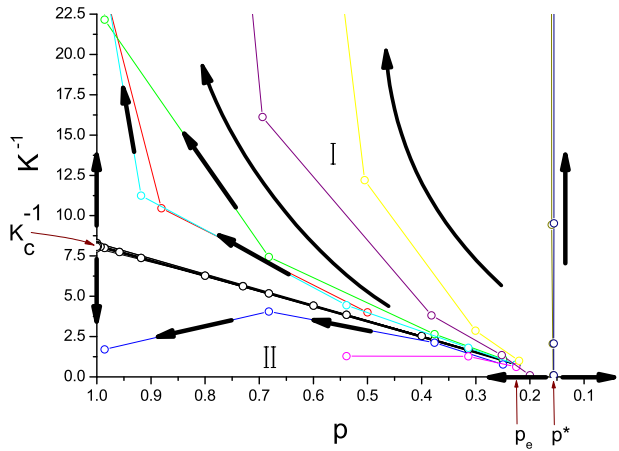}}}
\caption{The phase diagram, $K^{-1} = k_BT/J$ versus $p$, for
$d=3$. There is no fixed point other than the pure Ising one for
non zero temperatures. Thus the critical behaviour of the system
is determined by the pure Ising fixed point for finite $T$. Note
that the phase boundary comes down to zero temperature at a
concentration $p_e$ which is greater than the percolation fixed
point, $p^\ast$. (Our figure is in color in the on-line version.)}
\label{3d}
\end{figure}

For the bond diluted case, we first computed the RG flows for
$d=2$ and $d=3$ (See Figs. \ref{2d}, \ref{3d}). Due to the high
temperature approximation made in the determination of the RG
transformations, these flows are well defined on the disordered
side of the separatrix, and also for temperatures less than, but
close to, the transition temperatures, but not in the whole
ordered region. Nevertheless, their examination is crucial to
obtain the phase diagrams correctly.

For $d=2$, we find that the regions $\textrm{I}$ and $\textrm{II}$
flow, respectively, to the disordered and ordered fixed points at
$p=1$, $T \rightarrow \infty$ and $p=1$, $T=0$. The flow on the
separatrix itself is to the pure Ising fixed point indicated by
$T_{\rm c}$ on the $p=1$ line. Note that the line of fixed points
of the equation $a(\tilde p, K^{\ast}) = R(p, K^{\ast})$ extending
to the right of $(p^{\ast}, K^{\ast})$ and coming down to zero at
$p_{\rm e}$ is not a phase boundary, although it lies close to the
separatrix for $p\ge p^\ast$ and passes through an unstable fixed
point at $(p^{\ast}$, $K^{\ast})$. For $p< p^\ast$, we find that
in both regions $\textrm{III}$ and $\textrm{IV}$, the  flows are
to the attractive disordered fixed point at $p=0$, $T \rightarrow \infty$.
The line connecting the unstable fixed point at $p^{\ast}=0.5$,
$(K^{\ast})^{-1}=1.22$, indicated by $\ast$ in the figure, to the
percolation fixed point, $(p^\ast, 0)$ is therefore a first order
phase transition line, separating a region with finite
magnetization from one with zero magnetization. This suggests that
the unstable fixed point $(p^{\ast}, K^{\ast})$ is a tricritical
point (TCP), with a first order phase boundary connecting this
point to the percolation fixed point $p^\ast=0.5$ at $T=0$. We
have checked that along the separatrix, from the unstable disorder
fixed point to the pure Ising fixed point, the magnetization is
zero. (We have also checked that the mean field-type equation for
the equilibrium order parameter $m_0$, which one may obtain from
Eq. (\ref{corner0}) by setting all the magnetizations to be the
same, and interpreting the brackets as purely configuration
averages, gives a second order phase transition in this interval,
with the expected value of the order parameter critical exponent
$\beta =0.5$.)  The value of $\nu_{\rm dynamic}$ and $z$ at the
TCP are, $0.48$ and $2.16$ respectively. (Similar unexpected
features have arisen in other phase diagrams obtained via RSRG
treatments of systems with random bonds.~\cite{Falicov})

For $d=3$ the dynamical RG results for $\nu$ gives, via $\alpha =
2 - d \nu$, once again a positive value for $\alpha$ (although the
static result is negative, as can be readily computed from the
values in Table 1). However, we now find that there is no $K$
which the RG relation $a(p^\ast, K^{\ast}) = R(p^\ast, K^{\ast})$
is satisfied. Examining the flow diagram in Fig. \ref{3d}, we see
that the phase separation line comes down to $T=0$ at some $p_{\rm
e} > p^\ast$, precluding such a fixed point. The flow in regions
$\textrm{I}$ and $\textrm{II}$ are respectively to the disordered
and ordered fixed points, while on the separatrix it is once more
to the pure Ising fixed point.  For very low temperatures, near
$p_e$, the details of the phase boundary are not available, due to
the same difficulty as we encountered for $d=2$.

Since Monte Carlo simulations are plagued by crossover effects
along critical lines, we also computed effective critical
exponents $z_{\rm eff}$ along the critical line. For each given
$p$ along this line, we solved for a $p$-dependent fixed point of
$K$ under the transformation in Eq. (\ref{Kfix}) which now becomes
$a(f(p), K^\ast)=R(p, K^\ast)$. We then substitute $p$ and $K^\ast
(p)$ in Eq. (\ref{sfix}) and evaluate $z_{\rm eff}$ from ${\left(
\tilde s / s \right)}_{p, K^\ast} = \lambda^{z_{\rm eff}}$. We
find that for $d=2$, $z_{\rm eff}$ first increases from $2.13$ at
$p=1$ till $2.25$ at $p=0.75$ and then decreases to $2.01$ at
$p_e$. For $d=3$ the dependence on $p$ is again non-monotonic,
starting from $2.09$ at $p=1$, increasing to $2.69$ at $p=0.4$ and
decreasing to $2.54$ at $p_e$.

The calculation of the dynamical critical exponent $z$ for the
bond (or site) diluted quenched random Ising model are currently
the subject of numerical studies. Recent Monte Carlo simulations
for the dynamical critical exponent $z$ for the bond diluted
quenched random Ising model have only yielded an effective
exponent, $z_{\rm eff}$ , varying between  0.59 to 0.27 along the
critical line.~\cite{Berche} These values are markedly lower than the values
found here.

For $d \ge 4$, considering either the static or the dynamic values
of the correlation length exponent in Table 1, we find a negative
specific heat exponents from the hyper-scaling relation $\alpha = 2 - d\nu$.
This
suggests, although not conclusively~\cite{Janke}, that on all these lattices,
the pure Ising fixed point
will be attractive within this approach. Under dilution, the
critical behaviour on the second order phase boundary will be
determined by the pure Ising fixed point.  In Table 1 we also display the
dynamical critical exponent $z$ at the pure Ising fixed point for these values
of $d$.

It should be recalled that the exponents we have reported so far
are {\em exact} on hierarchical lattices~\cite{Berker,Kaufman}
generated by iterative replacement of each line in the cluster
shown in Fig. \ref{cluster}, by the cluster itself. We may define
an effective dimension of this hierarchical lattice as being given
via $2^{d_{\rm eff}-1}=N$, where $N$ is the number of intermediate
spins in the cluster in Fig. \ref{cluster}, and where we have
taken the scale transformation factor to be $\lambda = 2$. For
the undiluted case, we obtain  a network with a degree
distribution which is power law such that $n(\Gamma) \sim
{\Gamma}^{-\gamma}$, where $\gamma = 1 + \ln [2+(N-1)/2] / \ln N$.
We may now construct random hierarchical lattices by randomly
diluting each cluster, with a uniform bond occupation probability
$p$, to get the asymptotic degree distribution  $\sim \exp[-(1-p)\Gamma]/\Gamma!$ for small $p$. From the discussion above, we conclude that for effective dimension $d \ge 4$, the pure Ising
behaviour will be observed on the critical line for $T
> 0$, and $p > p_{\rm e}(d)$ on these hierarchical lattices.

{\bf  Acknowledgements}

AE would like to gratefully acknowledge partial support from the
Turkish Academy of Sciences.

\end{document}